\documentclass[%
reprint,
preprintnumbers,
 amsmath,amssymb,amsthm,superscriptaddress,
 aps,prd,
]{revtex4-2}

\usepackage{graphicx}% Include figure files
\usepackage{dcolumn}% Align table columns on decimal point
\usepackage{bm}% bold math
\usepackage{bbm}
\usepackage{subcaption}
\usepackage{amsthm}
\usepackage[usenames]{color}
\definecolor{PineGreen}{rgb}{0.0,0.47,0.44}
\definecolor{MidnightBlue}{rgb}{0.1,0.1,0.44}
\definecolor{magenta}{rgb}{1.0,0.0,1.0}
\usepackage{listings}
\definecolor{org1}{rgb}{.92,.39,.21}
\definecolor{pur2}{rgb}{.53,.47,.7}
\usepackage{pgfplots}
\usepackage{wrapfig}
\usepackage{float}
\pgfplotsset{compat=newest}
\usepackage{hyperref}
\usepackage{caption}
\newcommand{\orcid}[1]{\href{https://orcid.org/#1}{\includegraphics[width=10pt]{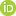}}}
\usepackage{footmisc}

\definecolor{myblue}{RGB}{72,127,227}
\definecolor{mygreen}{RGB}{48,142,48}
\definecolor{myviolet}{RGB}{57,49,103}
\definecolor{mymustard}{RGB}{213,147,62}

\definecolor{desycyan}{rgb}{0.00,0.68,0.93}
\definecolor{desyorange}{rgb}{0.96,0.52,0.07}
\definecolor{desygray}{rgb}{0.47,0.47,0.47}

\hypersetup{
	colorlinks=true,
	linkcolor=desycyan,
	citecolor=PineGreen,
	filecolor=magenta,
	urlcolor=desyorange
}

\newcommand{\Con}{\mathbf{Con}}

\newcommand{\RR}{\mathbb{R}}

\newcommand{\PP}{\mathbb{P}}

\newcommand{\pp}{\mathbb{P}}
\newcommand{\CC}{\mathbb{C}}
\newcommand{\ZZ}{\mathbb{Z}}

\newcommand{\Eu}{\mathbf{Eu}}

\newcommand{\bV}{\mathbf{V}}

\newcommand{\set}[1]{{\left\{{#1}\right\}}}

\newcommand{\cU}{\mathcal{U}}

\newcommand{\cF}{\mathcal{F}}

\newcommand{\cI}{\mathcal{I}}

\newcommand{\cM}{\mathcal{M}}

\newcommand{\Char}{\mathrm{Char}}
\newcommand{\mult}{\mathrm{mult}}

\newtheoremstyle{example}{}{}{}{}{\bfseries}{\smallskip}{\newline}{}
\theoremstyle{example}

\newcommand{\avg}[1]{\left< #1 \right>} % for average

\usepackage{tikz-feynman}
\tikzfeynmanset{compat=1.1.0}

\begin{document}

\title{Geometric Singularities of Feynman Integrals}%

\author{Martin Helmer\orcid{0000-0002-9170-8295}}
\email{{martin.helmer@swansea.ac.uk}}
\affiliation{Department of Mathematics, Swansea University, Swansea, Wales, UK.}

\author{Felix Tellander\orcid{0000-0001-6418-8047}}
\email{felix@tellander.se}
\affiliation{Mathematical Institute, University of Oxford, Oxford OX2 6GG, UK.}
\date{\today}

\begin{abstract}
We provide a new method to calculate the full microlocal description of singularities of Feynman integrals. This is done by associating a unique constructible function to the system of partial differential equations (PDEs) annihilating the integral and from this function the singularities can directly be read-off. This function can be constructed explicitly even if the system of PDEs is unknown and describes both the location of the singularities and the number of master integrals on them. Our framework is flexible enough to preform the calculation in any of the Lee-Pomeransky, Feynman, or momentum representations.
\end{abstract}

\maketitle
%\tableofcontents 
%\newpage
\section{Introduction}
Feynman integrals are an essential ingredient for calculations of observables in quantum field theory. Today their utility reaches far beyond their origin in high-energy physics and they are used for e.g.~calculations of gravitational waves and critical exponents in statistical field theory. Mathematically they strike the perfect balance of being concrete enough to allow for many explicit calculations to be carried out and general enough to have a rich and varied mathematical structure \cite{Weinzierl:2022eaz}.

The most basic invariant of any object is its singularities. For Feynman integrals this is a classical story \cite{Landau:1959fi}; culminating in the $S$-matrix program of the 60's \cite{Eden:1966dnq,Nakanishi1971}. Today there is vast renewed interest in studying these singularities \cite{Brown:2009ta,Pham2011integrals,Mizera:2021icv,Klausen:2021yrt,Caron-Huot:2024brh,Fevola:2023short,Helmer:2024wax,Hannesdottir:2024hke,Correia:2025yao,Matsubara-Heo:2025lrq}, with much focus placed on mathematical rigor and computability. In this paper we provide a new and simple way to describe these singularities as a constructible function. The pairing between a Feynman integral and this constructible function is not only rigorous but also explicitly computable on a computer.

The main results in the present paper are based on our recent article \cite{Helmer:2025yuf} and are given here without proof. Describing the necessary $D$-module theory for a rigorous proof of this construction distracts from its simplicity and is given in \cite{Helmer:2025yuf}. To emphasize the practicality of our approach we have implemented these methods in the \texttt{WhitneyStatifications} Macaulay2 \cite{M2} package which may be downloaded at the link below:
\begin{center}
    \url{http://martin-helmer.com/Software/WhitStrat/}.
\end{center}
\medskip

The paper is arranged as follows. We first provide some background on Whitney stratifications and related geometric objects. We then define constructible functions and characteristic cycles and show how they can be calculated for Feynman integrals. Our method is then applied to different representations of the Feynman integral.  

%%%%%%%%%%%%%%%%%%%%%%%%%%%%%%%%%%%%%%%%%%%%%%%%%%%%%%%%%%%%%%%%%%%%%%%%%%%%%%%%%%%%%%%%%%%%%%%%%%
\section{Geometric Preliminaries}
Before we define the main players in this paper; constructible functions and characteristic cycles, we introduce the objects necessary to both describe and calculate them. In the following, let $Y$ be a subvariety of dimension $d_Y$ in a manifold $X$. In this paper we take $X=\CC^n$.

\textbf{Whitney Stratification.} A fundamental tool for the study of singular spaces is a {Whitney stratification}. For a variety $Y$ this stratification divides it into smooth pieces which join together in a particularly desirable way. This allows for predictable behaviors of limiting tangent and secant lines between these pieces. More precisely, a {\em Whitney stratification} of $Y$ is a subdivision into a finite number of smooth disjoint connected manifolds $S_\alpha$, called {\em strata}, such that $Y=\sqcup_\alpha {S_\alpha}$ and such that {\em Whitney's condition} B holds for all pairs $M=S_\alpha$, $N=S_\beta$, of these manifolds. A pair of strata $(M,N)$, $M\subseteq\overline{N}$, satisfy Whitney's condition B \cite[Section 19]{Whitney1965} at a point $x\in M$ if for every sequence $\{p_\ell\} \subset M$ and $\{q_\ell\}\subset N$ with $\lim p_\ell=\lim q_\ell=x$, 
  the limit of secant lines between $p_\ell,q_\ell$ is contained in the limit of tangent planes to $N$ at $q_\ell$. When computing Whitney stratifications, and representing them on a computer, it is best to represent them as a flag $Y_\bullet$ of varieties $Y_0\subset \cdots \subset Y_{d_Y}=Y$ where the strata are the connected components of the successive differences $Y_i-Y_{i-1}$. 
  
\textbf{Conormal Variety.} Given a variety $Y$ in $\CC^n$ we may associate to it a conormal variety, ${\rm\bf Con}(Y)$, in $\CC^n\times \PP^{n-1}$ which catalogs all tangent planes and limiting tangent planes to each point $y$ in $Y$, in particular we also include singular points of $Y$ (in which case the planes cataloged are all limiting tangent planes). A point $(y, \eta)$ in ${\rm\bf Con}(Y)$ indicates that there is a tangent plane or limiting tangent plane to $Y$ at the point $y$ with normal vector $\eta$. More precisely, let $Y_{\rm Sing}$ denote the singular locus of $Y$ and let  $Y_{\rm reg} =Y-Y_{\rm Sing}$ denote the (open) manifold of smooth points in $Y$. The {\em conormal variety} is the subvariety in $Y\times \pp^{m-1}$ given by 
 \begin{align*}\label{eq:conormal}
{\rm\bf Con}(Y):=\overline{\left\lbrace  (y,\eta)\;|\; y \in Y_{\rm reg} \text{ and } T_yY_{\rm reg} \subset\eta^\perp \right\rbrace}.
\end{align*}

\textbf{Polar Variety.} While the conormal variety above catalogs all the tangents and limiting tangents to a variety, it also requires us to work in a larger ambient space. In many cases it is enough to understand which points in $Y$ have prescribed directions tangent to them, and in particular, to understand which points in $Y$ have tangent directions contained in a generically chosen linear space of some dimension. The object which catalogs these points in $Y$ with prescribed tangent directions is called a polar variety. Consider a flag $L_\bullet$ of length $d_Y+1$; $L_\bullet = \left(L_{d_Y+1} \supset L_{d_Y} \supset \cdots \supset L_{1} \right)$, where each $L_i \subset \CC^n$ is an $i$-dimensional linear subspace. For each $i$ in $\set{0,1,\ldots,d_Y}$, the dimension-$i$ \emph{polar variety} of $Y$ along the flag $L_\bullet$ is defined as the closure \small
  \[
  P_i(Y;\,L_\bullet) := \overline{\set{y \in Y_\text{reg} \mid \dim(T_yY_\text{reg} \cap L_{i+1}^\perp) > d_Y-i-1}}.
  \]\normalsize
  Note that $P_{d_Y}(Y;\,L_\bullet)=Y$.
  
\textbf{Euler Obstruction.} For a Whitney stratification $Y=\sqcup S_\alpha$, we define the \emph{Euler obstruction} of $\overline{S}_\gamma$ along $S_\alpha$, denoted $\Eu_{\overline{S}_\gamma}(S_\alpha)$, recursively as follows. We define $\Eu_{\overline{S}_\alpha}(S_\alpha)=1$ and for $S_\alpha\nsubseteq\overline{S}_\gamma$ we put $\Eu_{\overline{S}_\gamma}(S_\alpha)=0$. The remaining recursion is given by
\begin{equation}\label{eq: definition Euler obstruction}
\begin{split}
    \Eu_{\overline{S}_\gamma}(S_\alpha)&=\sum_{S\alpha\prec S_\beta\preceq S_\gamma}\Eu_{\overline{S}_\gamma}(S_\beta)c(S_\alpha,\,S_\beta),\\
    c(S_\alpha,\,S_\beta)&=\chi(B(x,\epsilon)\cap S_\beta\cap H_\eta),
\end{split}
\end{equation}
where $c(S_\alpha,\,S_\beta)$ is the topological Euler characteristic of a ball centered at $x\in S_\alpha$ with radius $\epsilon$ intersected with $S_\beta$ intersected with a generic linear affine space of codimension $\mathrm{dim}(S_\alpha)+1$ at a distance $\eta$ from $x$, where $\epsilon>>\eta>0$.

All the above constructions are closely related. Both conormal and polar varieties allow for algorithms to compute Whitney stratifications, see \cite{hnFOCM} resp.~\cite{helmer2023effective}. In fact, it has been shown by Teissier \cite{teissier1981varietes} that two strata $(M,N),\ M\subseteq\overline{N}$ satisfy Whitney's condition B if and only if the sequence of Hilbert-Samuel multiplicities \cite{eisenbud2013commutative}:
\begin{align*}
    &\mathrm{mult}_\bullet(\overline{N},z)=\\
    &\{\mathrm{mult}_z(P_0(\overline{N};\,L_\bullet)),\ldots,\mathrm{mult}_z(P_{d-1}(\overline{N};\,L_\bullet)),\mult_z(\overline{N})\}
\end{align*}
is constant for every $z\in M$. Given a Whitney stratification $\sqcup S_\alpha$ of $Y$, the above multiplicities are constant on strata and can therefore be indexed by the strata instead of $z$; that is for $z\in S_\alpha$ we write $\mult_\alpha(P_k(\overline{S}_\beta))$ for the constant value of  $\mult_z(P_k(\overline{S}_\beta))$. We can now use the multiplicities above to calculate the Euler obstruction \cite{BMM1994}:
\begin{equation}\label{eq: Euler obs from polar}
    \Eu_{\overline{S}_\beta}(S_\alpha)=\sum_{k=d_\alpha+1}^{d_\beta}(-1)^{d_\beta-k}\mult_\alpha(P_k(\overline{S}_\beta;\,L_\bullet)).
\end{equation}
%%%%%%%%%%%%%%%%%%%%%%%%%%%%%%%%%%%%%%%%%%%%%%%%%%%%%%%%%%%%%%%%%%%%%%%%%%%%%%%%%%%%%%%%%%%%%%%%%%
\section{Constructible Functions and Characteristic Cycles}
A \emph{constructible function} $\alpha$ on $X$ is a function of the form $\alpha(x)=\sum_j a_j\mathbbm{1}_{Y_j}(x)$ where $Y_j\subset X$ are subvarieties, $\mathbbm{1}_{Y_j}$ denotes the characteristic function (which equals $1$ for points in $Y_j$ and $0$ otherwise) and $a_j$ are integers. The Euler obstructions \eqref{eq: definition Euler obstruction} form another basis for the constructible functions, this is easily seen since $\Eu_{Y_j}(\cdot)$ generically equals one.  

Let $Q=\sum_\beta a_\beta(x)\partial^\beta$ denote a differential operator in local coordinates on $X$ of degree $m$ where $\beta$ is a multi-index. Then we define the \emph{principal symbol} to be $\sigma(Q)(x,\xi)=\sum_{|\beta|=m}a_\beta(x)\xi^\beta$. For a system of PDEs $\cM$ on $X$ given by $Q_1u=\cdots=Q_ru=0$ the \emph{characteristic variety} is given by\small
\begin{align*}
    &\Char(\cM)=\\
    &\{(x,\xi)\in X\times\PP^{n-1}\,|\,\sigma(Q)(x,\xi)=0\ \forall\ Q\in D_X\avg{Q_1,\ldots,Q_r}\}
\end{align*}\normalsize
where $D_X\avg{Q_1,\ldots,Q_r}$ denotes the left-ideal of differential operators in the Weyl algebra $D_X$ generated by $Q_1,\ldots,Q_r$. If $\Char(\cM)=\cup\Lambda_\alpha$ denotes the irreducible decomposition of $\Char(\cM)$, then every component is of the form $\Lambda_\alpha=\Con(Y_\alpha)$ for some subvariety $Y_\alpha\subset X$. The unique \emph{characteristic cycle} $CC(\cM)$ is now given by the formal sum $\sum_\alpha m_\alpha\Con(\overline{S}_\alpha),\ m_\alpha\in\ZZ$. This is an example of a \emph{conical Lagrangian cycle} \cite{kashiwara-schapira1}.

There is a simple isomorphism between the groups (or rather between the functors) of constructible functions and conical Lagrangian cycles:
\begin{equation}
    \Eu_{Y_\alpha}(\cdot)\simeq(-1)^{d_X-d_\alpha}\Con(Y_\alpha)
\end{equation}
where $d_X=\dim(X)$ and $d_\alpha=\dim(Y_\alpha)$. For a hypersurface $Y\subset X$ with Whitney stratification $\sqcup S_\alpha$ we get explicitly
\begin{equation}
    \begin{split}
             \mathbbm{1}_Y(y)&=\displaystyle\sum_\alpha(-1)^{d_Y-d_\alpha}m_\alpha\Eu_{\overline{S}_\alpha}(y)\quad\simeq\\
    CC(\mathbbm{1}_Y)&=-\displaystyle\sum_\alpha m_\alpha\Con(\overline{S}_\alpha).
    \end{split}
\end{equation}

It follows from Kashiwara's index theorem \cite{Kashiwara1973} that for a system of PDEs there exists a constructible function with the same characteristic cycle.  
%%%%%%%%%%%%%%%%%%%%%%%%%%%%%%%%%%%%%%%%%%%%%%%%%%%%%%%%%%%%%%%%%%%%%%%%%%%%%%%%%%%%%%%%%%%%%%%%%
\section{Geometric Singularities of Integrals}
We take the Lee-Pomeransky \cite{Lee2013} form as our representation of Feynman integrals when describing our construction;
\begin{equation}{\label{eq: Feynman integral}}    \cI=\int_{\RR_+^n}\left(\prod_{i=1}^n\frac{x_i^{\nu_i}\,dx_i}{x_i\Gamma(\nu_i)}\right)\,\frac{1}{(\cU(x)+\cF(x,z))^{D/2}}.
\end{equation}
However, as we will see in the next section this framework is flexible enough to deal with other representations as well.

To get all singularities we lift the integration variables $x$ to projective space using the Cheng-Wu theorem \cite{Weinzierl:2022eaz}. Let $Y\subset\PP_x^n\times\CC_z^m$ be the hypersurface defined by the vanishing of the homogenized integrand $x_0\cdots x_n(\cU x_0+\cF)=0$. As every factor in the integrand is raised to a parametric power, all of $Y$ is the singular locus for the system of PDEs annihilating the integrand. The rank of this system is one since the function defining $Y$ spans the solution space. So the constructible function that reproduces this system's characteristic variety is $\mathbbm{1}_{\PP^n\times\CC^m}-\mathbbm{1}_Y$. Expanding this we find
\begin{equation}\label{eq: char fnc annihilator}
    (\mathbbm{1}_{\PP^n\times\CC^m}-\mathbbm{1}_Y)(y)=1-\sum_{\alpha}(-1)^{d_Y-d_\alpha}m_\alpha \Eu_{\overline{S}_\alpha}(y)
\end{equation}
where $\sqcup S_\alpha$ is a Whitney stratification of $Y$. We note that we have all information necessary to calculate the integers $m_\alpha$. This is done by the following steps:
\begin{itemize}
    \item[(i)] Calculate a Whitney stratification of $Y$.
    \item[(ii)] Calculate the Euler obstructions using polar varieties \eqref{eq: Euler obs from polar}.
    \item[(iii)] Treat \eqref{eq: char fnc annihilator} as a linear system and solve for the $m_\alpha$.
\end{itemize}
If $\cM$ denotes the system of PDEs annihilating the integrand, then $\int \cM$ denotes the system annihilating the integral. For the characteristic variety we now have 
\begin{equation*}\small
 \Char(\cM)=\Con(0)\cup(\cup_i\Con(\overline{S}_{\alpha_i}))\subset\PP_x^n\times\CC_z^m\times\PP_\xi^n\times\PP_\zeta^{m-1}   
\end{equation*}\normalsize
 where $\alpha_i$ are the indices with non-zero $m_{\alpha_i}$ in \eqref{eq: char fnc annihilator}. Let $I_{\Char(\cM)}$ be an ideal in $\CC[x,z,\xi,\zeta]$ defining $\Char(\cM)$, then we define the elimination ideal 
\begin{equation}\label{eq: char ideal integral}
    I_{\Char(\int \cM)}=(I_{\Char(\cM)}+\avg{\xi_0,\ldots,\xi_n})\cap\CC[z,\zeta]
\end{equation}
whose variety is precisely the characteristic variety $\Char(\int \cM)$, see \cite{Helmer:2025yuf} and references therein.

We think of integration as being represented by the canonical projection map $\varphi: \pp_x^n\times \CC_z^m\to \CC_z^m $. The functorial properties of characteristic cycles and constructible functions means that we can obtain the constructible function associated to $\int \cM$ from $\cM$ by the proper push-forward:
\begin{equation}
    \varphi_*(\mathbbm{1}_V)(z)=\chi(\varphi^{-1}(z)\cap V)
\end{equation}
for any variety $V\subset\PP^n\times\CC^m$. Note that this assignment is unique \cite{MacPherson1974}. Specifically we obtain
\begin{widetext}
\begin{equation}\label{eq:varphiStar}
        \varphi_*(\mathbbm{1}_{\PP^n\times\CC^m}-\mathbbm{1}_Y)(z)=\chi((\PP^n\times\CC^m)\cap\varphi^{-1}(z))-\chi(Y\cap\varphi^{-1}(z))=n+1-\chi(Y\cap\varphi^{-1}(z))
\end{equation}
\end{widetext}
and we denote this function  in \eqref{eq:varphiStar} as $\chi_{\int \cM}(z)$. This is a constructible function on $\CC_z^m$ meaning that it has a decomposition into Euler obstructions. The singular locus $Z$ of $\int \cM$, i.e.~the \emph{Landau singularities} of the integral, is given by $Z=\overline{\pi_z(\Char^*(\int \cM))}$ where $\Char^*(\cdot)$ is the characteristic variety with the zero-section $\Con(0)$ removed and $\pi_z$ the canonical projection map $\CC_z^m\times\PP_\zeta^{m-1}\to\CC_z^m$. Given a Whitney stratification $\sqcup W_\alpha$ of $Z$, the constructible function is written as
\begin{equation}\label{eq: index theorem integral}\begin{split}
    \chi_{\int \cM}(z)&=n+1-\chi(Y\cap\varphi^{-1}(z))\\&=\mu_0-\sum_\alpha (-1)^{d_Z-d_\alpha}\mu_\alpha\Eu_{\overline{W}_\alpha}(z)
    \end{split}
\end{equation}
where $d_Z=\dim(Z)$, $d_\alpha=\dim(W_\alpha)$, and $\mu_i\in\ZZ$. The Euler characteristic $\chi(Y\cap\varphi^{-1}(z))$ is constant for $z\in W_\alpha$, hence this number is readily calculable using e.g.~\cite{helmer2016proj}. We therefore have all information needed to calculate the multiplicities $\mu_\alpha$ explicitly following the steps:
\begin{itemize}
    \item[(iv)] Calculate the singular locus $Z$ as the variety of the ideal $(I_{\Char(\int\cM)}:\avg{\zeta_1,\ldots,\zeta_m}^\infty)\cap\CC[z]$. This variety $Z$ is also the Landau singularities.
    \item[(v)] Whitney stratify $Z$ and calculate the Euler obstructions using polar varieties \eqref{eq: Euler obs from polar}.
    \item[(vi)] Calculate the Euler characteristic $\chi(Y\cap\varphi^{-1}(z))$ for a $z$ in each strata $W_\alpha$ and for a $z\in\CC_z^m-Z$, the latter equals $\mu_0$.
    \item[(vii)] Treat \eqref{eq: index theorem integral} as a linear system and solve for the $\mu_\alpha$.
\end{itemize}

Let $\alpha_i$ be the indices such that $\mu_{\alpha_i}\neq 0$, then we have the characteristic cycle
\begin{equation*}
    CC(\int \cM)=\mu_0\Con(0)+\sum_i\mu_{\alpha_i}\Con(\overline{W}_{\alpha_i}).
\end{equation*}

As we can determine the characteristic cycle directly from \eqref{eq: index theorem integral} and can write down this function for any Feynman integral, this function constitutes our main result. This function encodes significantly more than just the Landau singularities. First, the description of the singularities is in the cotangent bundle of $\CC_z^m$ so we get an upper bound on the set where solutions to the system $\int \cM$ fail to be microanalytic. Second, $\mu_0$ is the number of master integrals, i.e.~the dimension of the solution space. In fact, $\chi_{\int \cM}(z)$ is the dimension of the solution space at every $z$ so we also obtain the number of master integrals on singularities.

%%%%%%%%%%%%%%%%%%%%%%%%%%%%%%%%%%%%%%%%%%%%%%%%%%%%%%%%%%%%%%%%%%%%%%%%%%%%%%%%%%%%%%%%%%%%%%%%%%
\section{One Diagram, Three Integrals, Same Singularities}
To illustrate the method just described we calculate the singularities of the one-loop bubble using three different representations of the Feynman integral. Instead of using $z$ we denote the kinematic invariants as $\{p^2,\,m_1^2,\,m_2^2\}\in\CC^3$.
\begin{center}
\begin{tikzpicture}[baseline=-\the\dimexpr\fontdimen22\textfont2\relax,transform shape,scale=0.85]
    \begin{feynman}
    \vertex (a);
    \vertex [right = of a] (b);
    \vertex [right = of b] (c);
    \vertex [right = of c] (d);
    \diagram* {
	    (a) --[fermion] (b) -- [half left,edge label=\({x_1,\,m_1}\)](c) -- [anti fermion](d), (c) -- [half left,edge label'=\({x_2,\,m_2}\)](b),
};
    \end{feynman}
    \end{tikzpicture}
\end{center}
\textbf{Lee-Pomeransky.} In the Lee-Pomeransky representation we first have to homogenize the integrand. We have $Y=\bV(x_0x_1x_2(\cU x_0+\cF))\subset\PP_x^2\times \CC^3$ with the Symanzik polynomials $\cU=x_1+x_2$ and $\cF=(m_1^2+m_2^2-p^2)x_1x_2+m_1^2x_1^2+m_2^2x_2^2$. There are 24 strata in the minimal Whitney stratification of this surface and the seven strata of dimension zero and one appear with multiplicity zero in the characteristic cycle. Using \eqref{eq: char ideal integral} and eliminating the conormal variables we find the Landau singularities $Z=\bV(p^2m_1^2m_2^2((m_1^2+m_2^2-p^2)^2-4m_1^2m_2^2))$. The minimal Whitney stratification of this variety contains eleven strata where only the four codimension-one strata appear with non-zero multiplicity after solving \eqref{eq: index theorem integral}. Let $\lambda=\lambda(a,b,c)=(a+b-c)^2-4ab$ be the K\"all\'en function. The constructible function and characteristic cycle of the bubble integral is:
\begin{equation}\small
    \begin{split}
        \chi_{\int\cM}(z)=3&-\Eu_{\bV(p^2)}(z)-\Eu_{\bV(m_1^2)}(z)\\&-\Eu_{\bV(m_2^2)}(z)-(z)\Eu_{\bV(\lambda)}(z)\\
        CC(\int\cM)=\;&3\Con(0)+\Con(\bV(p^2))+\Con(\bV(m_1^2))\\
        &+\Con(\bV(m_2^2))+\Con(\bV(\lambda)).
    \end{split}
\end{equation}\normalsize
This shows that the bubble generically has three master integrals and that the number drops by one on each singularity.

\textbf{Feynman.} In the Feynman representation the integrand is already homogenized and given by the product of the Symanzik polynomials. For the bubble we have $Y=\bV(x_1x_2\cU\cF)\subset\PP_x^1\times\CC_z^3$. The Whitney stratification of $Y$ consists of 19 strata. All strata of dimension zero and one (in total seven strata) have multiplicity zero. The remaining strata have multiplicty one except $\bV(x_1,x_2)$ which has multiplicity four. Using \eqref{eq: char ideal integral} and eliminating the conormal variables we find the singular locus $Z=\bV(p^2m_1^2m_2^2((m_1^2+m_2^2-p^2)^2-4m_1^2m_2^2))$, the same as in the Lee-Pomeransky representation.

\textbf{Momentum.} The integrand in momentum representation is a product of propagators $q_i^2-m_i^2$ where $q_i$ is a linear combination of internal and external momenta. This is not homogeneous in the loop momenta, i.e. the integration variables, which will be denoted $k_i$ (with the homogenization variable denoted $k_h$) below. In momentum representation the number of integration variables and kinematic parameters depend on the intended number of space-time dimensions. For computational simplicity we assume that $m_1=m_2=m$. In two dimensions we have $Y_{2D}=\bV(k_h((k_0^2-k_1^2)-k_h^2m^2)((k_0+k_hp_0)^2-(k_1+k_hp_1)^2)-k_h^2m^2)\subset\PP_k^2\times\CC^3$ while in three dimensions we have $Y_{3D}\subset\PP_k^3\times\CC^4$ and similarly for higher dimensions. There are 18 strata in the minimal Whitney stratification of $Y_{2D}$ and the singular locus is $Z=\bV(m^2(p_0^2-p_1^2)(p_0^2-p_1^2-4m^2))$ which is the same as in the other representations with $m_1=m_2$.

%%%%%%%%%%%%%%%%%%%%%%%%%%%%%%%%%%%%%%%%%%%%%%%%%%%%%%%%%%%%%%%%%%%%%%%%%%%%%%%%%%%%%%%%%%%%%%%%%%
\section*{Acknowledgments}
FT is funded by the Royal Society grant number URF\textbackslash R1\textbackslash 201473. For the purpose of Open Access, the author has applied a CC BY public copyright licence to any Author Accepted Manuscript (AAM) version arising from this submission.

MH is supported by the Air Force Office of Scientific Research (AFOSR) under award
number FA9550-22-1-0462, managed by Dr.~Frederick Leve, and by the Royal Society under grant RSWF\textbackslash R2\textbackslash 242006, and would like to gratefully acknowledge this support.
%%%%%%%%%%%%%%%%%%%%%%%%%%%%%%%%%%%%%%%%%%%%%%%%%%%%%%%%%%%%%%%%%%%%%%%%%%%%%%%%%%%%%%%%%%%
\bibliographystyle{JHEP}
\bibliography{library}
\end{document}